\documentclass[conference]{IEEEtran}
\IEEEoverridecommandlockouts

\usepackage{cite}
\usepackage{amsmath,amssymb,amsfonts}
\usepackage{algorithmic}
\usepackage{graphicx}
\usepackage{textcomp}
\usepackage{xcolor}

\usepackage{amsmath}
\usepackage{xspace}
\usepackage{xcolor}
\usepackage{booktabs}
\usepackage{tikz}
\usepackage{changes}
\usepackage{booktabs}
\usepackage{graphicx}
\usepackage{url}

\def\BibTeX{{\rm B\kern-.05em{\sc i\kern-.025em b}\kern-.08em
    T\kern-.1667em\lower.7ex\hbox{E}\kern-.125emX}}

\usepackage{caption}
\graphicspath{ {./figures/} }

\usepackage{tikz}

\newcommand{\ra}[1]{\renewcommand{\arraystretch}{#1}}

\newcommand{\para }[1]{{\bf #1}}

\newif\ifdraft
\drafttrue

\ifdraft
\newcommand{\jbt}[1]{{\color{red}\textbf{JBT:}  \textit{#1}}}
\newcommand{\wfg}[1]{{\color{purple}\textbf{WFG:}  \textit{#1}}}
\newcommand{\tian}[1]{{\color{blue}\textbf{TG:}  \textit{#1}}}
\newcommand{\rjw}[1]{{\color{orange}\textbf{RJW:}  \textit{#1}}}
\else
\newcommand{\jbt}[1]{}
\newcommand{\tian}[1]{}
\newcommand{\rjw}[1]{}
\newcommand{\wfg}[1]{}
\fi

\newcommand{\1}{{\em (i)}}
\newcommand{\2}{{\em (ii)}}
\newcommand{\3}{{\em (iii)}}

\begin{document}

\date{}

\title{Memory-Efficient Deep Learning Inference in Trusted Execution Environments}

\author{
Jean-Baptiste Truong, William Gallagher\\
Tian Guo, Robert J. Walls\\
{\tt\small \{jtruong2, wfgallagher, tian, rjwalls\}@wpi.edu} \\
Worcester Polytechnic Institute\\
}

\pagestyle{plain}

\maketitle

\begin{abstract}
This study identifies and proposes techniques to alleviate two key bottlenecks
to executing deep neural networks in trusted execution environments (TEEs): page
thrashing during the execution of convolutional layers and the decryption of
large weight matrices in fully-connected layers. For the former, we propose a
novel partitioning scheme, \emph{y-plane partitioning}, designed to \1 provide
consistent execution time when the layer output is large compared to the TEE
secure memory; and \2 significantly reduce the memory footprint of convolutional
layers. For the latter, we leverage quantization and compression.  In our
evaluation, the proposed optimizations incurred latency overheads ranging from
1.09X to 2X baseline for a wide range of TEE sizes; in contrast, an unmodified
implementation incurred latencies of up to 26X when running inside of the TEE.

 \end{abstract}


\section{Introduction}

Deep learning model owners often rely on others' hardware, such as cloud
providers or end-users, for model execution. However, the consequent model
exposure impacts user privacy, model security, and the
owner's intellectual property interests~\cite{papernot2017practical,
shumailov2020sponge, brown2020language, dosovitskiy2020image}. To address these
issues, recent works have explored the use of \emph{trusted execution
environments (TEEs)}, i.e., isolated environments which provide a set of
security features that allow running verified code safely on untrusted
hardware~\cite{lee2019occlumency,kim2020vessels}. While TEEs provide a natural foundation for sensitive computations,
their severe memory constraints have important performance implications. In
the context of deep learning, where redundant access to large memory areas is
frequent, relying solely on existing TEE paging mechanisms results in
prohibitively high overheads---upwards of 26X increases in model latency (see
Table~\ref{tab:e2e_inference_latency}).

In this paper, we characterize two bottlenecks that impact TEE performance and
consider  methods to address them. For the \emph{page thrashing bottleneck}
described in Section~\ref{sec:yplane}, we propose a data partitioning scheme,
\emph{y-plane partitioning}, that allows for efficient computation of
convolutional layers in TEEs with as little as 28MB of secure memory.
Additionally, in Section~\ref{sec:connected}, we identify a previously
unexplored performance bottleneck, the \emph{decryption bottleneck}, that
arises from parameter decryption and propose a mitigation strategy based on
compression and quantization.  We used  SGX-based TEEs on Microsoft Azure cloud
servers~\cite{azure_sgx} to measure the impact of these bottlenecks and
evaluate the proposed solutions. For the most extreme case (shown in
Table~\ref{tab:e2e_inference_latency}), the bottlenecks increased model latency
to 26X over the unmodified baseline, while the proposed optimizations reduced
model latency from 26X to 1.09X.

The optimizations proposed in this study significantly reduce the per-layer
memory footprint for a model, which is a limiting factor of prior work such as
Vessels~\cite{kim2020vessels}. Further, we demonstrate that the proposed y-plane
partitioning scheme offers complimentary design tradeoffs, with different
strengths and weaknesses, to channel partitioning~\cite{lee2019occlumency}. Our
evaluation suggests that a combination of y-plane and channel partitioning
provides the smallest memory footprint for convolutional layers. The choice of
scheme depends on the size of the layer's output versus the size of the weights.
Finally, reducing memory footprint improves model latency and allows for greater
concurrency, allowing more TEEs to coexist on the same
system~\cite{kim2020vessels}. We leave an exploration of model concurrency for
future work. 

In summary, we make the following contributions:
\begin{itemize}
    \item the introduction of a novel \emph{y-plane partitioning} scheme that
      complements channel partitioning, alleviating the page thrashing
      bottleneck and reducing the memory footprint of convolution layers;    

    \item a characterization of the previously unexplored decryption bottleneck
    in fully-connected layers;

    \item an evaluation of quantization and compression as a means to address
      the decryption bottleneck and reduce the memory footprint of
      fully-connected layers.
\end{itemize}

 \section{Background}
\label{sec:motivation}

\begin{table}[t]
\centering
  \sffamily
  \ra{1.3}
  {\footnotesize
  \begin{tabular}{@{}lrrr@{}}
  \toprule
    \textbf{} &
    \textbf{} &
    \multicolumn{2}{c}{Inside TEE}\\ 
\cmidrule{3-4}
    \textbf{} &
  \textbf{Outside TEE} &
  \textbf{Optimized} &
  \textbf{Unmodified} \\
\textbf{} &
    \textbf{(s)} &
    \textbf{(s)} &
    \textbf{(s)} \\
  \midrule
  \textbf{28MB+1vCPU}  & 3.174    & 3.468 (1.09X)     & 84.639   (26.73X) \\
  \textbf{56MB+2vCPU}  & 1.858    & 2.430 (1.31X)       & 28.599 (15.39X)        \\
  \textbf{112MB+4vCPU} & 1.112    & 1.868   (1.68X)       & 11.256 (10.12X)        \\
  \textbf{168MB+8vCPU} & 0.808    & 1.667   (2.06X)        & 4.377 (05.42X)          \\
  \bottomrule
\end{tabular}} 
  \caption{\textbf{Model Latency of VGG-16.} 
  The ``Optimized'' column records the latency
  improvements after applying the optimizations proposed in
  Sections~\ref{sec:yplane} and~\ref{sec:connected}. Each row represents an SGX
  enclave configuration; for example, 28MB+1vCPU means the enclave has 28MB of
  secure memory and 1 virtual CPU core. Numbers are averaged over 30 runs.} 
\label{tab:e2e_inference_latency}
\end{table}

\para{Trusted Execution Environments.} While the
exact capabilities of TEE implementations vary, some of the more common security features include \1
isolation, i.e., the confidentiality and integrity of the code and data located
inside the TEE, and \2 remote attestation, i.e., the ability to verify the state of
the TEE remotely. These properties are why recent works have explored TEEs as a
means to protect the confidentiality of both user~\cite{lee2019occlumency} and
model~\cite{kim2020vessels} data when running deep learning inference on
untrusted hardware.

An \emph{SGX enclave} is a TEE implementation provided by Intel's software guard
extensions (SGX)~\cite{SGX}. SGX enclaves include an area of \emph{secure
memory}, called the \emph{processor reserved memory (PRM)}, which is isolated
from the rest of the system. This secure memory is only accessible from code
within the enclave.  The secure memory size is usually small relative to the
rest of the system, typically far less than what deep learning models require
for inference. For example, the enclaves used for this study offered between
28MB and 168MB of memory, whereas VGG-16~\cite{simonyan2014very} requires over
1GB of memory.  To support programs with higher
memory requirements, SGX provides paging mechanisms to encrypt and swap memory
pages between secure and main memory.  When code running inside the enclave
attempts to access a virtual memory address on a page that is not currently in
the enclave, a \emph{page fault} is raised. SGX transparently services this page
fault: evicting an older page and transferring, decrypting, and checking the
requested page's integrity. We refer to this as \emph{secure paging}.

The memory constraints for SGX-based TEEs differ from the memory constraints
for edge devices (e.g., mobile phones).  For example, the main memory size on
most mobile devices is much larger than the secure memory available to
the SGX enclaves used in our experiments. Further, SGX's secure paging
introduces unique encryption and decryption bottlenecks, as we discuss later. 

\para{Convolutional Neural Networks.} \emph{Convolutional neural networks
(CNNs)} are a type of deep neural network that contain neurons organized into
\emph{layers}, including the eponymous convolutional layers. CNNs are commonly
used for vision tasks but are garnishing attention in other domains.  The
process of using a CNN model for classification is  referred to as
\emph{inference} or \emph{model execution} and the time taken to perform this
inference is called \emph{model latency}.

\emph{Layer execution} refers to the process of transforming the \emph{inputs}
(i.e., the output of the previous layer) and \emph{parameters (e.g., weights)}
into the \emph{outputs} for an individual layer. The precise computation
performed depends on the layer's type. We collectively refer to the model's
static parameters (e.g., weights) and any values calculated at runtime as
\emph{model data}.

Broadly, CNNs use three different types of hidden layers:
\emph{fully-connected}, \emph{convolutional}, and \emph{pooling layers}. The
inputs and outputs of convolutional and pooling layers are 3D arrays which
resemble stacks of 2D images called \emph{channels} (e.g. the RGB channels of
an image). In fully-connected layers, the inputs and outputs are simple 1D
vectors.

 \section{Related Work}
\label{sec:related}

\para{Deep Learning and Trusted Execution Environments.}
In Vessels~\cite{kim2020vessels}, Kim et al. optimize the memory usage of
neural networks in TEEs by analyzing the dependency graph of the model's layers
and then allocating a  memory pool in which only the required data is stored at
any given time.  The rationale is that the sequential nature of neural
networks' architecture allows reusing most memory buffers, avoiding unnecessary
paging. Furthermore, as all of the computations are done in a pre-allocated
memory area, a single machine can host multiple enclaves to compute different
models concurrently. As long as the different enclaves do not fill up the
secure memory, the contention is minimized. The limiting
factor for such a system is the size of the memory pool, which relies
primarily on the size of the largest layers.

Partitioning is one mechanism to reduce the per-layer memory requirements.  For
example, we propose a convolution-layer partitioning scheme, y-plane
partitioning, in Section~\ref{sec:yplane}.  Another example is Occlumency's
\emph{channel partitioning}.  Occlumency~\cite{lee2019occlumency} is an
inference framework implemented on top of SGX that uses channel partitioning to
divide the computation and memory requirements of convolutional layers. 

As our work does not propose an end-to-end system, we cannot provide a
direct comparison with Occlumency and Vessels. However, we compare Occlumency's
channel partitioning scheme to y-plane partitioning and explain why a
combination of both schemes provides the best performance in
Section~\ref{subsec:large-layers}.

Grover et al.\ proposed Privado~\cite{tople2018privado}, a system designed to
remove any input-dependent memory accesses, thereby preventing information
leakage from the TEE. Chiron~\cite{hunt2018chiron} uses SGX enclaves to train
machine learning models, protecting the confidentiality of the user's training
data, the model's architecture, and the training procedure.  Neither work
attempts  to address the performance challenges described in this paper.

\para{Encryption for Deep Learning.} Cryptographic techniques offer an
alternative to trusted execution environments for maintaining user
privacy~\cite{gilad2016cryptonets, mishra2020delphi}. These techniques rely on
homomorphic encryption to process encrypted data on a server. Such systems
usually have high inference latencies, which they make up for with high
throughput. Thus, these systems are more appropriate for processing large batches of
input data. Further, existing cryptographic systems like
CryptoNets~\cite{gilad2016cryptonets} do not protect model weights from
disclosure---protecting the model confidentiality in Cryptonets would
significantly degrade performance.

 \section{Methodology}
\label{sec:methodology}

We conducted our experiments on virtual machines provided by Microsoft's Azure
cloud computing infrastructure. The four tested enclave configurations
represent all of the configurations offered by Microsoft Azure at the time of
writing.  We refer to each VM using its enclave size and number of virtual
CPUs; for example, \emph{28MB+1vCPU} refers to the VM configuration with 28MB of
secure memory and 1 virtual CPU. All configurations ran Ubuntu 18.04 and used
an Intel Xeon E-2288G CPU. Unless otherwise specified, the number of execution
threads for each system was equal to the number of virtual CPUs---this is why
the baseline model latency varies, for example.

Our evaluation methodology emphasizes the \emph{per-layer} performance of
convolutional neural networks (CNNs).  Focusing on individual layers offers two
distinct benefits.  First, it allows us to examine each of the components in
isolation. Second, it helps us determine the performance implications for a
variety of CNN architectures.  For instance, we observed that the performance
benefits offered by quantization and compression for the large fully-connected
layers in VGG-16 directly translated to performance benefits for the large
fully-connected layer in AlexNet---though we elide the AlexNet numbers for
space. Consequently, we focus primarily on the VGG architecture as VGG models
contain a variety of fully-connected and convolutional layers that range in
size, shape, and memory requirements.

We use Darknet as the baseline inference framework due to its
portability. In particular, Darknet uses C and lacks external dependencies,
making it possible to port the framework to SGX with relatively minor changes.
In contrast, PyTorch is a more popular framework, but moving model execution
entirely into the TEE would require significant engineering efforts. 
 \section{The Page Thrashing Bottleneck}
\label{sec:yplane}

The mismatch between enclave size and convolutional-layer memory requirements
manifests as inefficient paging patterns during model inference, i.e., a
\emph{page thrashing bottleneck}. In this section, we characterize this
phenomenon.  We then propose y-plane partitioning as a means to mitigate this
bottleneck. We compare y-plane partitioning to a prior scheme and argue that
the combination  offers the best latency and smallest memory footprint.

\subsection{Characterization} 

As observed by Lee et al.~\cite{lee2019occlumency}, the challenge for
convolutional layers is that the memory access pattern during execution leads to
page thrashing, i.e., the constant transfer of pages into and out of the
TEE.\footnote{Denning~\cite{denning68thrashing} defines thrashing as ``excessive
overhead and severe performance degradation or collapse caused by too much
paging.''} Every transfer between secure and main memory adds significant
overhead.

Darknet, like many other frameworks, uses the im2col transformation to speed up convolutional layer
execution. This transformation expands the 3D input array into a large 2D matrix
and organizes the weights into a different 2D matrix. These transformations
allow the convolution operations to be computed using a large matrix-matrix
multiplication. They thus benefit from highly optimized \emph{general matrix to
matrix multiplication (GEMM)} functions provided by BLAS
libraries (e.g. OpenBLAS~\cite{xianyi2012openblas}). This method's inherent tradeoff is that the
im2col transformation duplicates the inputs, resulting in a transformed im2col
matrix that is significantly larger than the original input array.

The above scheme has unintended consequences when used naively in the TEE.
First, im2col's expansion of the input array---a factor of 9 in VGG-16---causes
many memory pages to be evicted from the TEE only to be brought back into the
TEE during the matrix-matrix multiplication.  
Second, as the im2col matrix cannot fit entirely in TEE memory, the pattern of
memory accesses to this matrix has important performance implications.
For example, in Darknet the output matrix is computed row by row, resulting in an
unfavorable memory access pattern that triggers cascading evictions and page faults.
In particular, computing one row of the
output requires a lookup of the entire transformed im2col matrix, and this
lookup process is repeated for all rows of the output. 

In our experiments with Darknet and VGG-16, we observed that
convolutional layers cause more page evictions, by multiple orders of magnitude,
when run in a 28MB enclave versus the 168MB enclave. Layer 8 triggers 1.8
million page evictions in the 28MB enclave, but only 1,700 in the 168MB enclave.

\subsection{Partitioning}

Partitioning addresses the thrashing bottleneck by applying the im2col
transformation to a subset of the input array. The subset, i.e., partition, can
be processed efficiently using the limited secure memory of the TEE.

\begin{figure}
  \centering
   \includegraphics[width=\columnwidth]{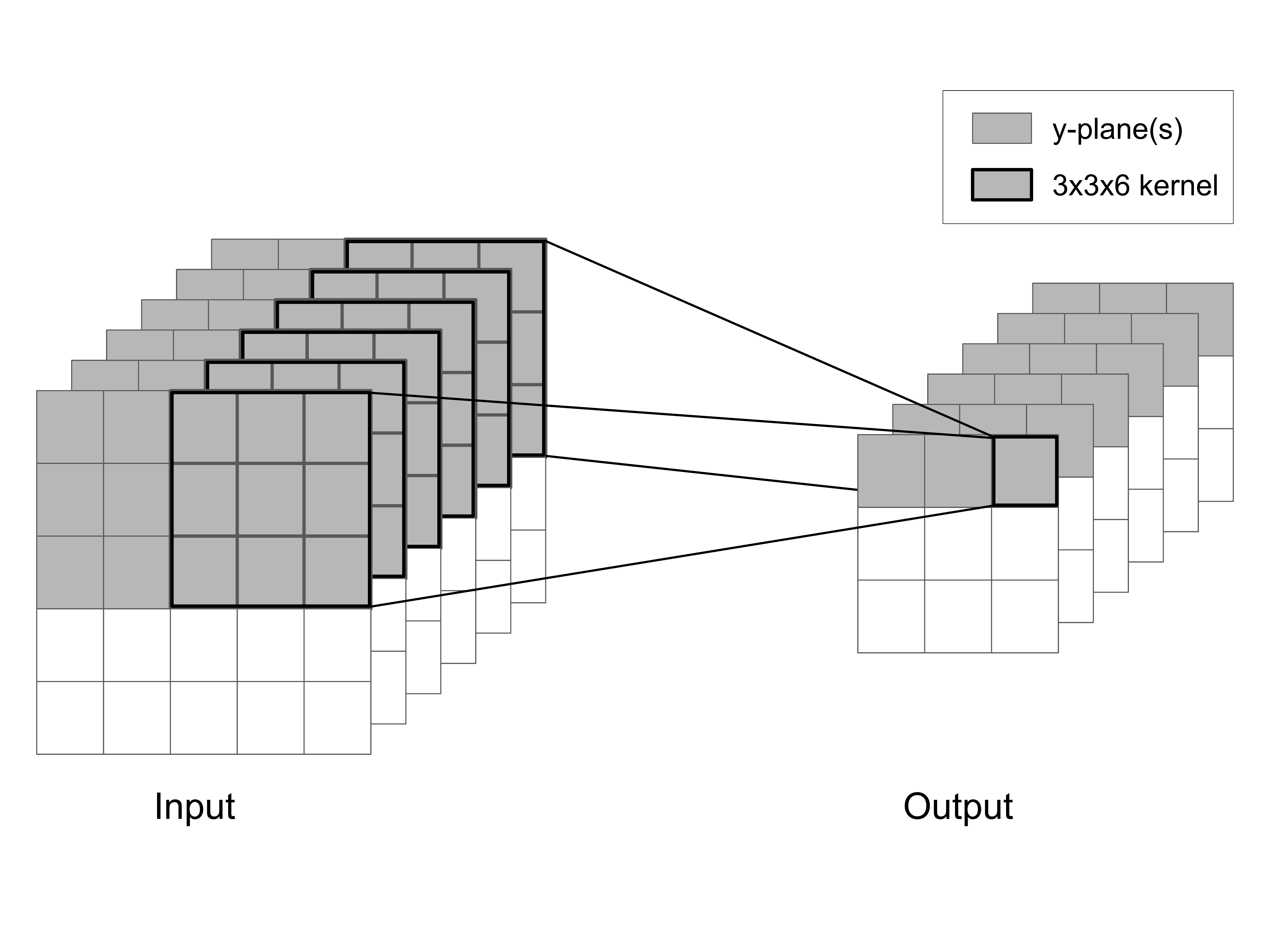}
   \caption{\textbf{Illustration of Y-Plane Partitioning.} A 5x5x6 input is convolved with
   a 3x3x6 kernel. This figure highlights the computation of 1 output value.
   Three input y-planes are required to compute one output y-plane.}
   \label{fig:y-planes-channels}
 \end{figure}

\begin{figure}
 \centering
  \includegraphics[width=\columnwidth]{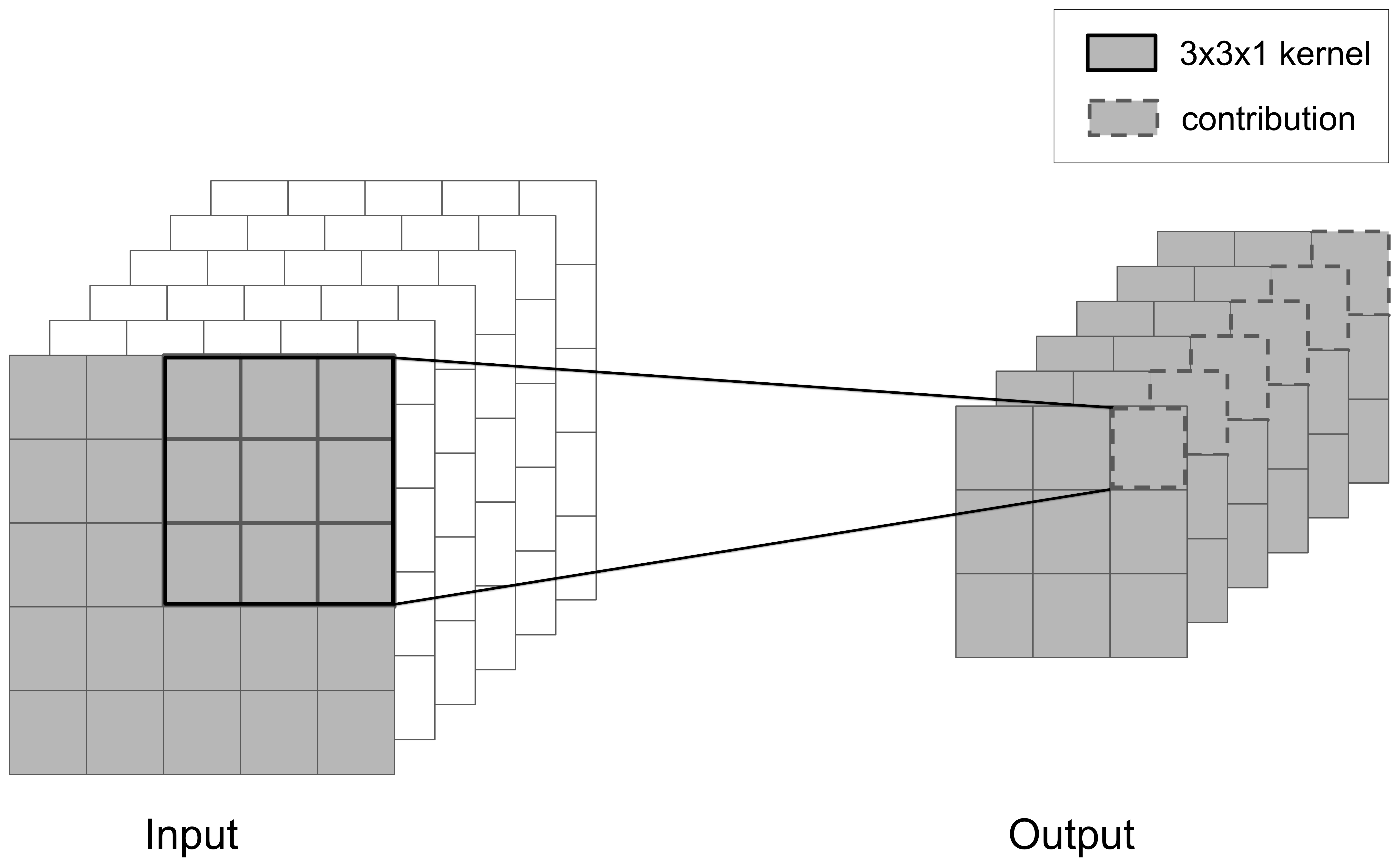}
  \caption{\textbf{Illustration of Channel Partitioning.} This figure highlights
  the contribution of 1 input channel to 1 value on each output channel. Each
  input channel contributes to the entire output.}
  \label{fig:channels-partitioning}
\end{figure}

We evaluate two partitioning schemes: \emph{y-plane partitioning} and
\emph{channel partitioning}. The former is a contribution of this paper, and the
latter was previously used as part of Occlumency~\cite{lee2019occlumency}.
While channel partitioning splits the input by channels, \emph{orthogonal} to
the depth direction, \textit{y-plane partitioning} uses planes \emph{parallel}
to the depth direction; Figure~\ref{fig:y-planes-channels} provides a visual
representation of y-plane partitioning.

\begin{table*}[t]
  \centering

{\footnotesize
  \sffamily
\begin{tabular}{@{}lrrrrrrrrrrrrr@{}}
  \toprule
  \textbf{} &
  \textbf{} &
  \phantom{} &
  \multicolumn{2}{c}{\textbf{28MB+1vCPU}} & 
  \phantom{} &
  \multicolumn{2}{c}{\textbf{56MB+2vCPU}} & 
  \phantom{} &
  \multicolumn{2}{c}{\textbf{112MB+4vCPU}} & 
  \phantom{} &
  \multicolumn{2}{c}{\textbf{168MB+8vCPU}}\\ 
  \cmidrule{4-5}  
  \cmidrule{7-8}  
  \cmidrule{10-11}  
  \cmidrule{13-14}  
  \textbf{Layer} &
  \textbf{Input} &&
  \textbf{Latency} &
  \textbf{Evictions} &&
  \textbf{Latency} &
  \textbf{Evictions} &&
  \textbf{Latency} &
  \textbf{Evictions} &&
  \textbf{Latency} &
  \textbf{Evictions} \\
  \textbf{} &
    \textbf{(MB)} &&
  \textbf{(s)} &
  \textbf{(\#)} &&
  \textbf{(s)} &
  \textbf{(\#)} &&
  \textbf{(s)} &
  \textbf{(\#)} &&
  \textbf{(s)} &
  \textbf{(\#)} \\
  \midrule
    \textbf{Outside TEE}\\
    \qquad 2  & 123   && 0.547 & - && 0.340 & - &&
    0.219 & - && 0.168 & -\\
    \qquad 5  & 61  && 0.455 & - && 0.274 & - &&
    0.152 & - && 0.106 & -\\
    \qquad 8  & 31  && 0.416 & - && 0.237 & - &&
    0.127 & - && 0.085 & -\\
    \textbf{Partitioning in TEE}\\
    \qquad 2  & 123  && 0.399 & 8,270 && 0.223 & 3,503 &&
    0.162 & 3,251 && 0.153 & 3,184\\
    \qquad 5  & 61  && 0.296 & 1,999  && 0.177 & 1,716 &&
    0.120 & 1,718 && 0.101 & 1,712\\
    \qquad 8  & 31  && 0.292 & 1,374 && 0.172 & 1,364 &&
    0.102 & 1,362 && 0.084 & 1,360\\
    \textbf{Unmodified in TEE}\\
    \qquad 2  & 123 && 17.899 & \textbf{1,859,815} && 10.989 &
    \textbf{1,881,017} &&
    6.153 & \textbf{1,858,259} && 0.746 & 30,286\\
    \qquad 5  & 61  && 17.664 & \textbf{1,838,563} && 10.651 &
    \textbf{1,848,772} &&
    0.682 & 8,799 && 0.381 & 1,729\\
    \qquad 8  & 31  && 17.556 & \textbf{1,827,409} && 1.121 & 4,909 &&
    0.538 & 1,361 && 0.323 & 1,700\\

  \bottomrule
  \end{tabular}
} \caption{\label{tab:conv_thrashing_bottleneck}\textbf{Latency and Page
Evictions for Convolution Layers using Y-plane Partitioning.} Only the three
convolutional layers with the largest inputs are shown. Input size was measured
after the im2col transformation.
}
\end{table*}

At a high level, both schemes first split the input into partitions and compute
the contribution of that partition to the output by \1 applying im2col on each
partition, \2 computing a matrix-matrix multiplication with the corresponding
subset of the weight matrix, and \3 adding the result to the output buffer.
Y-plane and channel partitioning offer different design tradeoffs, with
complementary strengths and weaknesses. In particular, we find that y-plane
partitioning is more memory-efficient when the layer output is large, while
channel partitioning is better when the weight matrix is large.

\para{Y-Plane Partitioning.}
As illustrated in Figure~\ref{fig:y-planes-channels}, \emph{y-planes} are the
concatenation of one row from each channel of a 3D array. 
For this scheme, a partition is a group of contiguous y-planes; both the
layer's inputs and outputs are logically divided into y-plane partitions.  Each
output y-plane is computed from a small and contiguous subset of the input
y-planes.  The convolution kernel size and stride determine the relationship
between input and output y-planes.

Each round of computation involves three elements: \1 an output partition composed
of contiguous y-planes, \2 the corresponding subset of input y-planes, and \3
the entire weight array. This repeated access to the entire weight array makes
the weights size the limiting factor of y-plane partitioning.

\para{Channel Partitioning.}
Channel partitioning, illustrated in Figure~\ref{fig:channels-partitioning},
divides the input into partitions of one or more
channels~\cite{lee2019occlumency}, using a partition of the weights to calculate
each contribution to the output.  Note that the output is not partitioned and
needs to be accessed during every round of computation to add the input-weight
partition pairs' contribution.  Thus, the output size is the limiting factor. 

In practice, deep neural networks contain many convolutional layers, and the
output and weight sizes of each layer vary. This
observation, along with the aforementioned differences between y-plane and
channel partitioning, suggests that a combination could yield the best results.
Such a scheme would use the best partitioning scheme for the given layer.
Further, the cost of switching from y-plane to channel partitioning (and vice
versa) is negligible.  We explore this idea in
Section~\ref{subsec:large-layers}.

\subsection{Performance of Y-plane Partitioning}

Table~\ref{tab:conv_thrashing_bottleneck} illustrates the page thrashing
bottleneck in convolutional layers, showing the impact of enclave size on
latency and page evictions for the unmodified baseline running outside of the
TEE, inside of the TEE, and inside of the TEE with
y-plane partitioning. We make several observations that are consistent with
prior work~\cite{lee2019occlumency}.

First, with y-plane partitioning, the convolutional layer latency decreased
significantly and remained stable for all secure memory sizes.
Second, page thrashing in Darknet was triggered when the size of the
im2col-transformed input exceeded the enclave size---as measured by the
drastic difference in page evictions. For example, layer 2, with its 123MB
input, saw approximately 1.8 million page evictions for all three enclaves with
less than 123MB of secure memory, but only 30 thousand evictions for the
enclave with 168MB of memory. 

Third, Darknet's per-layer latency varied dramatically, ranging from more than
17 seconds when thrashing occurred to less than 1 second when thrashing did not
occur. As the total number of floating point operations remained constant, this
difference resulted from thrashing.

\subsection{Combining Y-Plane and Channel Partitioning}
\label{subsec:large-layers}

Different factors limit Y-Plane and Channel partitioning. Below we demonstrate
those differences using a model with layer sizes that far exceed the available
secure memory. We show that a combination of y-plane and channel partitioning
allows us to execute this model without thrashing, whereas either scheme would
fail if used in isolation.

\para{Methodology.} 
To scale up the model,
we preserved most layer parameters (stride, padding, etc.), types, structure,
and order of VGG-16. We only scale up two parameters: \1 the input resolution,
which has an impact on the input and output size in all the layers, and \2 the
number of kernels in the first layer, which impacts the inputs, outputs, and
weights size in all the layers. We chose an input resolution of $450\times 450$
and $64$ kernels in the first layer, so that some layers have either their
output or weights larger than the enclave size. We call this model
\emph{VGG-Large}.

\begin{table}[]
    \begin{tabular}{@{}rrrrrrr@{}}
    \toprule
    \multicolumn{1}{c}{\textbf{}} & \multicolumn{1}{c}{\textbf{}} & \multicolumn{1}{c}{\textbf{}} & \multicolumn{2}{c}{\textbf{Y-Plane}} & \multicolumn{2}{c}{\textbf{Channel}} \\ \cmidrule(l){4-7} 
    \multicolumn{1}{l}{}     & Weights                       & Output                        & Latency       & Evictions            & Latency     & Evictions              \\
    \multicolumn{1}{l}{}    & (MB)                          & (MB)                          & (s)           & (\#)                 & (s)         & (\#)                   \\ \midrule
    1                             & 0.01                          & \textbf{49.44}                & 0.544         & 48,883               & 1.387       & \textbf{123,145}       \\
    2                             & 0.14                          & \textbf{49.44}                & 1.877         & 62,964               & 21.552      & \textbf{1,809,386}     \\
    4                             & 0.28                          & \textbf{24.72}                & 0.891         & 22,001               & 9.216       & \textbf{868,540}       \\
    5                             & 0.56                          & \textbf{24.72}                & 1.638         & 29,444               & 18.204      & \textbf{1,716,377}     \\
    7                             & 1.13                          & 12.47                         & 0.770         & 5,310                & 1.205       & 5,301                  \\
    8                             & 2.25                          & 12.47                         & 1.542         & 10,283               & 2.392       & 7,680                  \\
    10                            & 4.50                          & 1.64                          & 0.288         & 1,606                & 0.166       & 1,579                  \\
    11                            & 9.00                          & 1.64                          & 0.583         & 2,916                & 0.330       & 2,727                  \\
    13                            & 9.00                          & 0.44                          & 0.239         & 2,434                & 0.107       & 2,420                  \\
    14                            & 9.00                          & 0.44                          & 0.237         & 2,419                & 0.107       & 2,422                  \\
    16                            & 17.58                         & 0.06                          & 0.112         & 4,543                & 0.066       & 4,602                  \\
    17                            & \textbf{34.33}                & 0.06                          & 0.467         & \textbf{35,689}      & 0.126       & 8,901                  \\
    18                            & \textbf{68.66}                & 0.12                          & 0.925         & \textbf{70,979}      & 0.252       & 17,885                 \\
    19                            & \textbf{68.66}                & 0.06                          & 0.925         & \textbf{70,878}      & 0.253       & 17,762                 \\ \bottomrule
    \end{tabular}
    \caption{\label{tab:large_model}\textbf{Per-Layer Latency and Page Evictions for
    VGG-Large.}}    
    \end{table} 
\para{Results.} Table~\ref{tab:large_model} shows the per-layer page evictions
and inference latency for y-plane partitioning and channel partitioning.  We
make four observations of these results.  First, when  the output is large
compared to the secure memory size, as is the case in the first few layers,
channel partitioning will start thrashing. In contrast, y-plane partitioning
divides the output and, consequently, saw up to 58X fewer page evictions than
channel partitioning. 

Second, in the last few layers the weights are larger than secure memory, and
y-plane partitioning shows up to 4.0X more page evictions than channel
partitioning.  This behavior is expected as y-plane partitioning does not divide
the weights, but channel partitioning does. 

Third, each scheme out-performed the other for a subset of the layers.  In other
words, using y-plane and channel partitioning in conjunction allows for
efficient computations for models that neither y-plane nor channel partitioning
could handle without page thrashing. To completely avoid page thrashing with
VGG-Large, an enclave of at least 68 MB (resp. 50 MB) would be needed to run this model with
y-plane-only (resp. channel-only) partitioning; while the hybrid scheme can run it
with just 28MB. 
This experiment also shows that a large model can be ran with a significantly
reduced memory footprint even if it can fit in memory. This result is useful in
practice as systems that provide concurrency for secure deep learning inference,
like Vessels~\cite{kim2020vessels}, are limited by the memory footprint of
individual models. Therefore, this hybrid scheme is likely to allow for greater
concurrency, enabling more models to share the available secure memory
efficiently. Of course, our observations are incomplete, and it is essential to
consider other factors, such as the specifics of the target model and other
potential sources of concurrency-based contention. We leave such explorations
for future work. 

Lastly, for the layers that can fit both the output and weights in secure
memory, channel and y-plane partitioning are comparable in terms of latency and
number of page evictions. Further experiments showed that, for these
intermediate layers, the slight difference between both schemes is due to the
GEMM (matrix multiplication) implementation. When using standard  GEMM libraries
such as OpenBLAS~\cite{xianyi2012openblas}, this difference disappeared. Thus, we
do not claim that one scheme is superior to the other for layers that fit in
secure memory.
 \section{The Decryption Bottleneck}
\label{sec:connected}

\begin{table*}[t]
  \centering
  {\footnotesize
  \sffamily
  \begin{tabular}{@{}lrrrrrrrrrrrrr@{}}
  \toprule
  \textbf{} &
  \textbf{} &
  \phantom{} &
  \multicolumn{2}{c}{\textbf{28MB+1vCPU}} & 
  \phantom{} &
  \multicolumn{2}{c}{\textbf{56MB+2vCPU}} & 
  \phantom{} &
  \multicolumn{2}{c}{\textbf{112MB+4vCPU}} & 
  \phantom{} &
  \multicolumn{2}{c}{\textbf{168MB+8vCPU}}\\ 
  \cmidrule{4-5}  
  \cmidrule{7-8}  
  \cmidrule{10-11}  
  \cmidrule{13-14}  
  \textbf{Layer} &
  \textbf{Input} &&
  \textbf{Latency} &
  \textbf{Evictions} &&
  \textbf{Latency} &
  \textbf{Evictions} &&
  \textbf{Latency} &
  \textbf{Evictions} &&
  \textbf{Latency} &
  \textbf{Evictions} \\
  \textbf{} &
    \textbf{(MB)} &&
  \textbf{(s)} &
  \textbf{(\#)} &&
  \textbf{(s)} &
  \textbf{(\#)} &&
  \textbf{(s)} &
  \textbf{(\#)} &&
  \textbf{(s)} &
  \textbf{(\#)} \\
  \midrule
    \textbf{Outside TEE}\\
    \qquad 19  & 392 && 0.030 & - && 0.029 & - &&
    0.030 & - && 0.029 & -\\
    \qquad 21  & 64  && 0.005 & - && 0.005 & - &&
    0.005 & - && 0.005 & -\\
    \qquad 23  & 16  && 0.001 & - && 0.001 & - &&
    0.001 & - && 0.001 & -\\
    \textbf{Quant./Comp. in TEE}\\
    \qquad 19  & 392  && 0.542 & 55,707 && 0.498 & 55,258 &&
    0.497 & 16,167 && 0.464 & 16,168\\
    \qquad 21  & 64  && 0.090 & 9,194  && 0.094 & 11,157 &&
    0.076 & 2,660 && 0.075 & 2,660\\
    \qquad 23  & 16  && 0.023 & 2,090 && 0.020 & 2,500 &&
    0.020 & 715 && 0.022 & 715\\
    \textbf{Unmodified in TEE}\\
    \qquad 19  & 392  && 0.987 & 101,113 && 1.158 & 102,328 &&
    1.213 & 101,270 && 1.124 & 103,536\\
    \qquad 21  & 64  && 0.162 & 16,481 && 0.191 & 16,516 &&
    0.203 & 16,458 && 0.185 & 16,523\\
    \qquad 23  & 16  && 0.039 & 4,090 && 0.046 & 4,165 &&
    0.050 & 4,030 && 0.045 & 4,041\\
  \bottomrule
  \end{tabular}}\caption{\label{tab:decryption_fc} \textbf{Latency for Fully-Connected Layers
using Quantization and Compression.}
Numbers averaged over 30 runs.}
\end{table*}

 \begin{table}[t]
  \centering
  {\footnotesize
  \ra{1.3}
  \sffamily
  \begin{tabular}{@{}lrrrrrrrrrrr@{}}
  \toprule
  \textbf{} & & & \multicolumn{5}{c}{\textbf{Compression Ratio}} \\ 
  \cmidrule{4-8}
  \textbf{Accuracy} &
  \textbf{Base.} &
  \textbf{Quant.} &
  \multicolumn{1}{c}{\textbf{32:10}} &
  \multicolumn{1}{c}{\textbf{32:5}} &
  \multicolumn{1}{c}{\textbf{32:4}} &
  \multicolumn{1}{c}{\textbf{32:3}} &
  \multicolumn{1}{c}{\textbf{32:2}} \\ 
  \midrule
  \textbf{Top-1 (\%)} & 70.4 & 70.4 &  70.4 &  70.2 &  68.1  & 68.5    &  26.7     \\
  \textbf{Top-5 (\%)} & 89.8 & 89.8 &  89.8 &  89.8 &  89.1  & 89.0    &  53.6    \\ 
  \bottomrule
  \end{tabular}}
  \caption{\label{tab:fc_accuracy}\textbf{Model Accuracy with Quantization and
  Compression.} 
  A compression ratio of 32:10 means that a buffer of 32 bytes is compressed
  into 10 bytes. We omit ratios from 32:9 to 32:6 as they produced the same
  results as ratio 32:10.} 
\end{table}
 
Partitioning alleviates the page thrashing bottleneck. Without thrashing, the
transfer of model parameters into the enclave becomes the dominating
performance factor due to the overhead of page decryption and integrity
checking. This issue, which we call the \emph{decryption bottleneck}, is
especially problematic for fully-connected layers with large weight matrices.
We explore quantization and compression as possible solutions, reducing the
number of pages that need to be transferred. 

\subsection{Characterization}

For ease of exposition, we refer to the collection of components that handle
secure paging---i.e., the eviction, encryption/decryption, and integrity checking
of pages---as the \emph{decryption link}. 

In fully-connected layers, loading the weights into secure memory is expensive.
For example, in our experiments, we observed that the first fully-connected
layer of VGG-16 (i.e., layer 19) took 0.028 seconds to execute with Darknet
normally, but 1.131 seconds ($\sim$40X) to execute with Darknet when run inside
a trusted execution environment. Our experiments show that the difference in
execution time was due entirely to the additional 1.102 seconds needed for
loading in the weights from main memory---the 0.028 seconds needed for the layer
computations was trivial by comparison. 

Assuming we cannot modify the hardware to improve secure paging performance, and
because the decryption link is already saturated, we turn toward techniques to
\emph{reduce the amount of data} that must be transferred over that link.
Specifically, we analyze the use of two techniques, quantization
and compression, to reduce the size of the weights for fully-connected layers.
Further, as these techniques require additional computation, multi-threading can
be used to keep the decryption link saturated.   

Quantization is the process of converting the set of
possible weight values (e.g., 32-bit floats) into a smaller discrete set of
values (e.g., 16-bit floats). Some information is lost in this conversion,
potentially affecting the model's accuracy, but the total memory
requirements are halved. The weights are stored quantized and are converted
back to 32-bit floats once decrypted. The cost of converting the values back to
32-bits floats was negligible in our experiments.

Similarly, compression also reduces data transfer requirements.  We only
consider lossy compression here as the compression factor for lossless
compression was too small to be useful in our experiments.  The amount of
information lost is directly related to the compression factor, which can be
tuned for many compression algorithms. The computational cost of decompression
is higher than quantization, but the workload can be split more easily between
virtual CPUs.

\subsection{Performance of Quantization and Compression}
\label{subsec:eval-quant-comp}

Table~\ref{tab:decryption_fc} shows the execution latency for fully-connected
layers. For the 28MB+1vCPU and 56MB+2vCPU enclave configurations, we observe
roughly  half as many page evictions as unmodified Darknet, and execution took roughly half of the time.  This
performance difference is due to the quantization scheme, which halves the size
of the weight matrix.  In separate experiments, we observed that adding more
than two threads failed to yield further improvement for quantization,
suggesting that two threads are sufficient to saturate the decryption link.

Compression benefits more than quantization from the larger number of virtual
CPUs offered by the 112MB+4vCPU and 168MB+8vCPU configurations.  When using
compression, the number of page evictions decreased to roughly 16\% of
unmodified Darknet.  Once the decryption link was saturated with 6 threads,
the compression scheme proved more efficient than quantization in these
enclaves.

Lastly, we observe no drop in accuracy from quantization, as shown in
Table~\ref{tab:fc_accuracy}. Results will vary by model, and the impact of
quantization on accuracy is an active area of research in the AI
community~\cite{krishnamoorthi2018quantizing, zhou2017incremental,
dundar1995effects}. More aggressive quantization strategies could yield even
higher performance. For compression in all but the most extreme compression
rate, the top-1 accuracy was within 2\% of baseline and the top-5 accuracy was
within 0.8\%.

 \section{Conclusions}
\label{sec:conclusions}

In summary, we studied the use of partitioning, quantization, and compression
to improve the memory efficiency of deep learning inference in trusted
execution environments. Partitioning addresses the page thrashing bottleneck,
and a combination of the proposed y-plane partitioning scheme and channel
partitioning allows for the smallest memory footprint.  Quantization and
compression reduce the impact of the decryption bottleneck with little impact
on model accuracy.

The primary limitation of this study is the limited number of models we
consider. While convolution and fully-connected layers are common to a wide range of
deep learning models, the benefits of the aforementioned optimizations depend
on model specifics. For example, partitioning will not reduce the inference
latency for the layers that already fit in memory (e.g. in ResNet).  Even so, partitioning
allows for a configurable  memory footprint. This configurability is especially
important in the context of concurrent inference, i.e., multiple enclaves
running on a single server. We believe a full study of partitioning and model
concurrency is an interesting direction for future work.

It is also important to consider other hardware capabilities when configuring
the optimizations, such as secure memory size and decryption speed. For
example, one would need to adjust the size of each partition to ensure they fit
within secure memory. In the current implementation, a partition can be as
small as a single y-plane, which for VGG-16 is at most a few hundred
kilobytes. As another example, one might also want to tune the compression and
quantization factors based on the decryption speed and available CPU resources.

Finally, TEE-based model inference is a building block for more complex
security and privacy guarantees. For example, we could extend TEE-based
inference to protect user privacy by hiding the user's input from both the
hardware owner and the model provider. Supporting this feature would require
additional components, such as a secure communication channel between the TEE
and the user to hide both the user's input and the inference results from the
hardware provider. Again, we leave such efforts for future work.

\bibliographystyle{IEEEtranS}
\bibliography{./bib/bib}

\end{document}